\documentclass[twocolumn,showpacs,prl]{revtex4}
\usepackage{amsmath}
\usepackage{amsfonts}
\usepackage{amssymb}
\usepackage[english]{babel}
\usepackage{graphicx}
\usepackage{bm}

\begin{document}

\title{Abrikosov vortex escape from a columnar defect \\
as a topological electronic  transition in vortex core}
\author{A.~S.~Mel'nikov and A.~V.~Samokhvalov}
\affiliation{Institute for Physics of Microstructures, Russian
Academy of Sciences,  603950 Nizhny Novgorod, GSP-105, Russia}

\date{\today}

\begin{abstract}
We study microscopic scenario of vortex escape from a columnar
defect under the influence of a transport current. For defect
radii smaller than the superconducting coherence length the
depinning process is shown to be a consequence of two subsequent
topological electronic transitions in a trapped vortex core. The
first transition at a critical current $j_L$ is associated with
the opening of Fermi surface segments corresponding to the
creation of a vortex--antivortex pair bound to the defect. The
second transition at a certain current $j_d > j_L$ is caused by
merging of different Fermi surface segments, which accompanies the
formation of a freely moving vortex.
\end{abstract}


\maketitle
The study of  magnetic and transport properties of type-II
superconductors in the presence of artificial pinning centers is
known to be an important direction in the physics of vortex matter
\cite{Blatter-RMP94}. An obvious first step to understanding of
the pinning related phenomena is associated with the evaluation of
the individual pinning force acting on a vortex in the presence of
a defect. Various approaches used for such analysis are based on
the observation that a defect attracts the vortex in order to
avoid loss of condensation energy in the core and to decrease a
kinetic energy due to perturbations of screening currents and
magnetic field induced by the defect
\cite{Mkrtchyan-JETP72,Buzdin-PhysC96,%
Buzdin-PhysC98,Nordborg-PRB00}.
Close to the critical temperature
$T_c$ both these effects can be described within the
 Ginzburg--Landau (GL) theory valid for large length
scales well exceeding the superconducting coherence length $\xi$
at zero temperature.

Such long wavelength approach is, of course, no more valid for low
temperatures $T \ll T_c$ and small defect size $a\lesssim\xi$.
Microscopic approach for calculation of the pinning energy and
appropriate modification of the GL functional for a particular
case of a small point impurity with the scattering cross section
$\sigma_{sc} \ll \xi^2$ have been previously studied in Ref.~
\cite{Thuneberg,Thuneberg-JLTP84}. Considering the microscopic
theory one should take into account the behavior of the subgap
fermionic states bound to the vortex core which are known to
determine both the structure and dynamics of vortex lines in the
low temperature limit. These subgap states are known to form the
so--called anomalous spectral branch crossing the Fermi level. For
well separated vortices the behavior of the anomalous branches can
be described by the Caroli--de Gennes--Matricon (CdGM) theory
\cite{Caroli-PL64}: for each individual vortex the energy
$\varepsilon_0(\mu)$ of subgap states varies from $-\Delta_0$ to
$+\Delta_0$ as one changes the angular momentum $\mu$ defined with
respect to the vortex axis. Here $\Delta_0$ is the superconducting
gap value far from the vortex axis. At small energies
$|\varepsilon|\ll\Delta_0$ the spectrum is a linear function of
$\mu$: $\varepsilon_0(\mu,k_\perp) \simeq -\mu\hbar\omega$, where
$\hbar\omega\approx\Delta_0\ln\Lambda/(k_\perp\xi)$ is the
interlevel spacing, $\xi=\hbar V_F/\Delta_0$,
$k_\perp^2=k_F^2-k_z^2$, $k_z$ is the momentum projection on the
vortex axis,  $k_F$ and $V_F$ are the Fermi momentum and velocity,
respectively. The logarithmic factor with $\Lambda\sim\Delta_0/T$
appears at low temperatures due to the so--called Kramer--Pesch
effect \cite{Kramer-Pesch,Kopnin-TNSC}.
  Neglecting for $T \gg
\hbar\omega$ the quantization of the angular momentum $\mu$ one
can consider the anomalous branch to cross the Fermi level at
$\mu=0$ for all orientations of the Fermi momentum $\mathbf{k}_F$.
Thus, in the space $\mu - \mathbf{k}_F$ we obtain a Fermi surface
(FS) for excitations localized within the vortex core (see
\cite{Volovik-Universe} for review). Changing magnetic field or
transport current we can get switching between different vortex
states which is accompanied by the changes in the FS topology.
Such topological transitions in quasiparticle spectra of vortex
systems are similar to the well--known Lifshits transitions which
occur in the band spectra of metals \cite{lifshits,blanter}. One
can separate two generic examples of such transitions in vortex
matter: (i)~opening of FS segments corresponding to the creation
of vortex-antivortex pairs \cite{Volovik-gapless,Melnikov-PRB08};
(ii)~merging and reconnection of different FS segments via the
Landau-Zener tunneling \cite{mel-silaev,Melnikov-PRB08}. The basic
properties of vortex matter such as pinning and transport
characteristics, heat transport in the vortex state and
peculiarities of the local density of states should be strongly
affected by these changes in the FS topology. The goal of this
Letter is to suggest a theory of topological electronic
transitions which occur in a pinned vortex core under the
influence of transport current.

The mechanism of these transitions is closely related to the
effect of elastic scattering on the CdGM levels. Resulting
modification of the anomalous spectral branch is noticeable even
for impurity atoms introduced in a vortex core \cite{Larkin-PRB98}
and becomes much more pronounced provided we consider defects of
the size well exceeding the Fermi wavelength. The absence of the
FS is a hallmark of a vortex pinned by an insulating columnar
defect and it is natural to expect that vortex depinning should be
accompanied by the FS formation. For a vortex line trapped by such
defect the subgap spectrum has been analyzed recently within the
quasiclassical approach in Ref.~\cite{Melnikov-PRB09D}. For rather
large angular momenta $|\mu|>\mu_a=k_\perp a$ quasiclassical
trajectories of electron--hole excitations do not experience
reflection at the defect surface and, thus, the subgap spectrum
coincides with the CdGM one. In the opposite case $|\mu|\leq\mu_a$
the normal reflection of trajectories at the defect destroys the
low energy part of CdGM spectral branch: with the decrease in
$|\mu|$ below the threshold value $\mu_a$ the energy of the subgap
bound state rapidly approaches the superconducting gap $\Delta_0$
(see Fig.~1). Thus, contrary to the CdGM case the subgap spectral
branch of the pinned vortex does not cross the Fermi level: there
appears a minigap $\Delta_m (a)  \sim |\varepsilon_0 (k_Fa,k_F)|$
in the quasiparticle spectrum. The spectral branch $\varepsilon_a$
for trajectories touching the defect can be approximated by
vertical lines  passing from $\pm\Delta_m$ to $\pm\Delta_0$ for
$\mu=\mp\mu_a$ (see Fig.~1b).
%
\begin{figure}
\begin{center}
\includegraphics[width=0.35\textwidth]{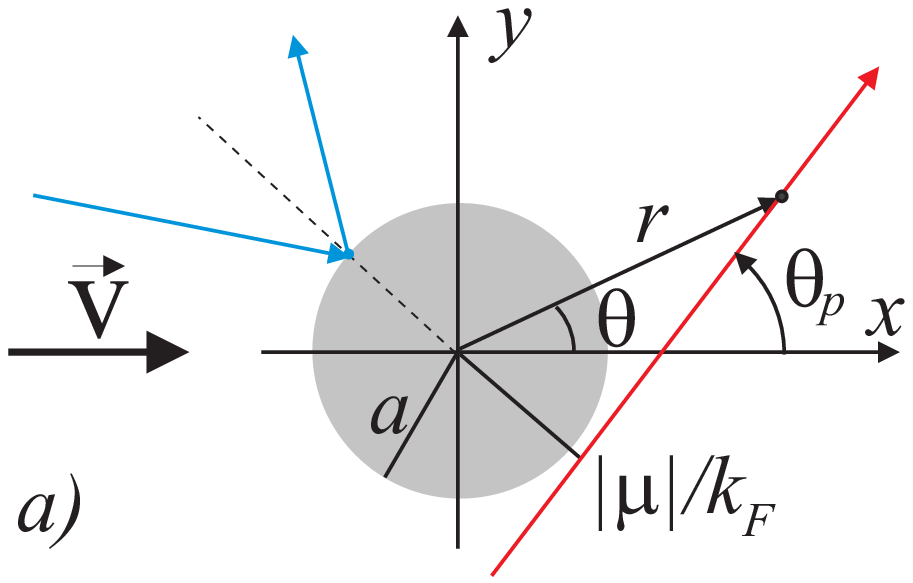}
\includegraphics[width=0.35\textwidth]{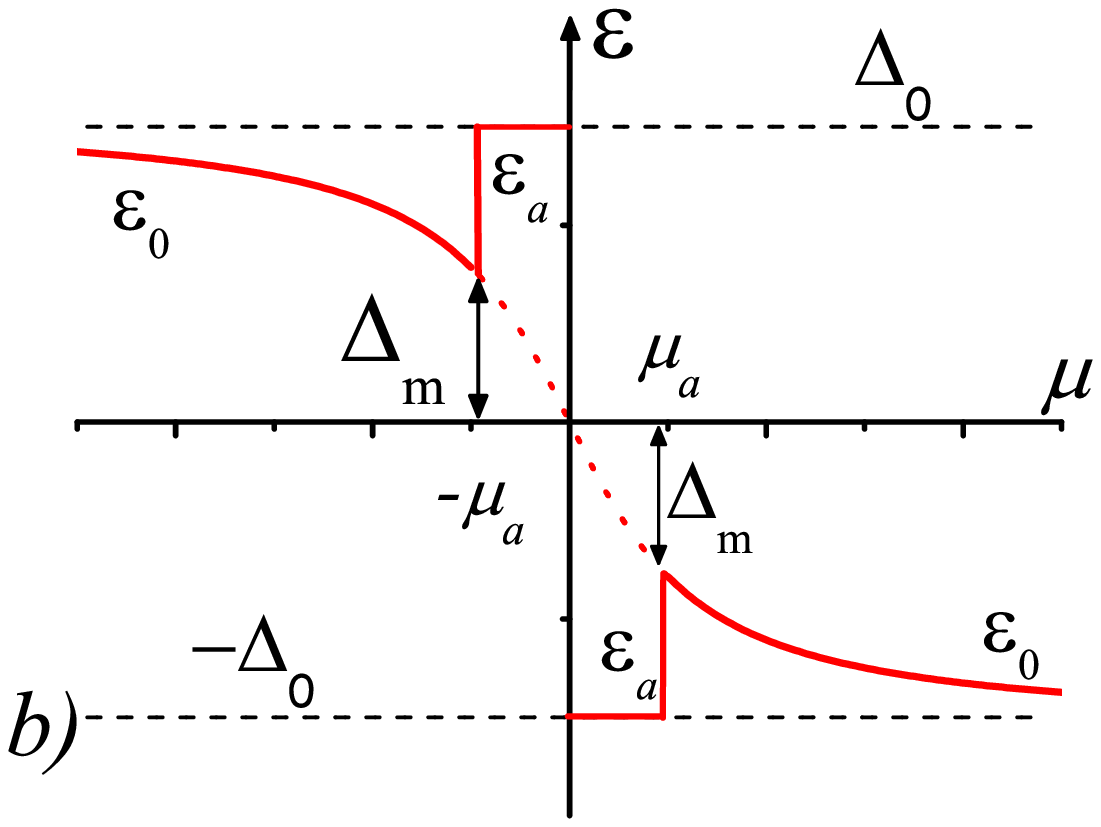}
\end{center}
\caption{\textbf{Fig.1} Different types of quasiclassical
trajectories (a) and spectrum of subgap states (b) for a vortex
pinned by the defect.} \label{Fig:1}
\end{figure}
%

 In
this Letter we propose a microscopic scenario of vortex escape
from a columnar defect of a radius $a$ smaller than the
 coherence length $\xi$ under the influence of an
external supercurrent $\mathbf{j} = e n_s \mathbf{V}$ applied
perpendicular to the vortex axis (Fig.1). The transport
supercurrent results in a Doppler shift $\varepsilon_d =
(\hbar\mathbf{k}\cdot\mathbf{V})$ of the quasiparticle energy and
the subgap spectrum  takes the form:
\begin{equation}\label{eq:2}
\tilde{\varepsilon}(\mu,k_\perp,\theta_p) = \varepsilon
(\mu,k_\perp) + \hbar k_\perp V\cos\theta_p \ ,
\end{equation}
where $\varepsilon=-\Delta_0\,{\rm sign}\,\mu$ for quasiparticles
with $|\mu|<\mu_a$ and $\varepsilon=\varepsilon_0$ for
quasiparticles with $|\mu|>\mu_a$.
 Here the quasiparticle momentum $\mathbf{k}=k_\perp ( \cos
\theta_p,\,\sin \theta_p)$ defines the trajectory orientation
angle $\theta_p$ in the ($x,\,y$) plane and $\mathbf{V}=V
\mathbf{x}_0$. The depinning process can be viewed as two
subsequent topological electronic transitions in the vortex core
which restore the anomalous spectral branch crossing the Fermi
level and repair the FS. The first transition at the critical
current $j_L = e n_s V_L$ is associated with the opening of
separate segments of the FS which appear when the Doppler--shifted
branch of the quasiparticle spectrum crosses the Fermi level and
the minigap in the quasiparticle spectrum vanishes:
\begin{equation}
V \gtrsim V_L = \Delta_m(a) / \hbar k_F \ .
\end{equation}
The latter condition coincides, in fact, with the famous Landau
criterion of superfluidity \cite{LL9}.
  The corresponding critical current density
can be expressed via the depairing current density ${j_c = e
k_F^2\Delta_0 / 3\pi^2\hbar}$ and CdGM spectrum:
\begin{equation}\label{eq:3}
    j_L = j_c\, |\varepsilon_0(k_Fa,k_F)|/\Delta_0 \leq j_c \ .
\end{equation}
Considering the zero temperature limit and assuming $a \ll \xi$ we
obtain the following expressions for the critical velocity and
current:
\begin{equation}
V_L  = \Delta_0 a \ln\Lambda/ \hbar k_F \xi \ ,\quad
 j_L = j_c (a/\xi)\ln\Lambda \ll j_c \ ,
\end{equation}
where $\Lambda\sim\xi/a$. Taking $\tilde \varepsilon =0$ in
Eq.(\ref{eq:2}) we find the equation defining FS segments in the
($\mu,\,\theta_p$) plane:
\begin{equation}
\label{orbitt}
\varepsilon_0 (\mu,k_\perp) =- \hbar k_\perp V\cos\theta_p \ ,
\quad |\mu|>\mu_a \ .
\end{equation}
These segments are shown in Fig.~2 by solid lines in the limit
$a\ll\xi$ when $\varepsilon_0 (\mu,k_\perp)=-\hbar\omega\mu$. The
ends of the segments are joined by the lines $\mu=\pm\mu_a$ which
correspond to the spectral branch $\varepsilon_a$. The direction
of quasiparticle trajectory precession along the resulting closed
classical orbits in the $\mu - \theta_p$ plane is determined by
the Hamilton equation: $\hbar\,
\partial\theta_p / \partial t =
\partial \tilde{\varepsilon} / \partial \mu$.
The formation of FS segments is obviously associated with the
appearance of superconducting phase singularities outside the
columnar defect.
%
\begin{figure}[t!]
\begin{center}
\includegraphics[width=0.4\textwidth]{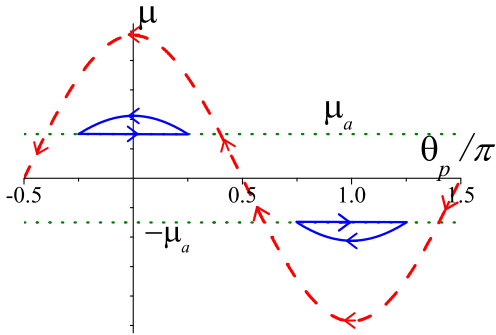}
\end{center}
\caption{\textbf{Fig.2} Schematic plot of Fermi surface segments
in the ($\mu, \theta_p$) plane for  $V_L < V \ll V_d$ (solid blue
line) and $V > V_d$ (dashed red line).} \label{Fig:2}
\end{figure}
%

We now proceed with the analysis of the structure of a resulting
vortex state bound to the defect.
 For the sake of simplicity hereafter we put $k_\perp=k_F$ which
 is appropriate for strongly anisotropic layered material with
 vortex lines oriented along the anisotropy axis.
 One can separate two contributions to the gap function
 $\Delta(r,\theta)$ in the self-consistency equation:
 (i)~the term $\Delta_{1}$ associated with the subgap states localized
 within the vortex core; (ii)~the term $\Delta_{2}$
 which is defined by the quasiparticle states with large energies
 $\varepsilon\gtrsim\Delta_0$. We introduce here the polar
 coordinate system $(r,\theta)$ in the plane perpendicular to the
 vortex axis and the origin chosen at the cylindrical defect
 center (see Fig.~1).
 At rather low temperatures the  term $\Delta_2$ is only weakly
 affected by the transport current $j\ll j_c$ since the
 corresponding Doppler shift is much smaller than the
 superconducting gap $\Delta_0$.
On the contrary, the first contribution to the gap is strongly
modified in the presence of a transport current provided the
Doppler shift exceeds the minigap $\Delta_m (a)$.
 Our further calculations of the current induced correction to the
  term $\Delta_{1}(r,\theta)$  are based on
a semiclassical version of the self-consistency equation (see
\cite{Volovik-JETPL93}):
\begin{equation}\label{eq:6}
    \delta\Delta_1(r,\theta)  =
    \frac{\Delta_b}{4 i} \int\limits_0^{2\pi} d \theta_p\, e^{i \theta_p}\,
        \left({\rm sign}\,\tilde{\varepsilon} -
        {\rm sign}\,\varepsilon\right)\,,
\end{equation}
where  $\varepsilon(\mu)$ ($\tilde\varepsilon (\mu,\theta_p)$) is
the subgap spectrum in the absence (presence) of transport
current, $\mu=k_Fr\sin(\theta_p-\theta)$,  and $\Delta_b \sim
\Delta_0$ is the amplitude of the gap value contribution coming
from the CdGM spectral branch.
 Here we neglect the coordinate dependence of normalized electron- and hole-like
wavefunctions at the scale $r \sim a \ll \xi$ and consider the
zero temperature limit. Close to the threshold velocity $V_L$ the
spectrum $\tilde\varepsilon$ can be considered perturbatively in
the form (\ref{eq:2}) with $k_\perp=k_F$.
 Note that here we restrict ourselves to the small $\mu$ limit assuming $j\ll j_c$.
 For small superfluid velocities $V \le V_L$, the gap
function $\Delta_1 = \Delta_b e^{i\theta}$ is undisturbed
($\delta\Delta_{1}=0$). Otherwise, if $V
> V_L$, the Doppler shift results in  qualitative changes in the
gap function distribution. A resulting piecewise regular gap
function
$$\Delta_1(r,\theta)=|\Delta_1| \exp(i \phi)=\Delta_b\,
\mathrm{e}^{i \theta}+\delta\Delta_1(r,\theta)$$ is defined in a
set of regions shown in Fig.~3:
\begin{equation}\label{eq:8}
    \Delta_1(r, \theta) = \Delta_b  \mathrm{e}^{i \theta}
        \begin{cases}
            1, & \text{in $A$ or $C$}\,, \\
             1 - 2 \sqrt{r^2-a^2} /r, & \text{in $B$}\,, \\
           1 - \mathrm{e}^{i \alpha_r} + \mathrm{e}^{-i \alpha_0} , & \text{in $D$}\,, \\
             1 - \mathrm{e}^{-i \alpha_r} + \mathrm{e}^{-i \alpha_0}, & \text{in $E$}\,,
        \end{cases}
\end{equation}
$$ {\rm where}\,\,\, \sin \alpha_r = \frac{a}{r}\,, \,\,\, \sin
\alpha_0 = \frac{r_0 \cos \theta}{ \sqrt{r^2+2 r_0 r \sin \theta +
r_0^2}}\ , $$ and $r_0= V /\omega$.  By dashed lines in Fig.~3 we
also show schematically the distribution of the superconducting
phase gradient $\nabla\phi$ around the defect. One can clearly
observe strong changes in topology of the phase $\phi$
distribution caused by the superflow: there appear a phase
singularity at the point $S$ ($x=0,\, y=-r_0$) and the line $L$
($r = 2 a / \sqrt{3}$) of zeros of the superconducting order
parameter $\Delta_1$ in the region $B$ (see Fig.3). The
singularity at the point $S$ has the same vorticity as the initial
vortex captured by the defect. With an increase in the superfiuid
velocity $V$ this singularity moves away from the defect. One can
see that the modification of the phase $\phi$ distribution shown
in Fig.~3 corresponds to the formation of a bound
vortex-antivortex pair. Note that the vortex and antivortex
positions found here coincide with the positions of the phase
singularities of a total gap function since the amplitude of the
second part of the gap function $\Delta_2$ is rather small:
$|\Delta_2(r\sim r_0)|\sim \Delta_0 r_0/\xi\ll \Delta_b$.
%
\begin{figure}[t!]
\begin{center}
\includegraphics[width=0.3\textwidth]{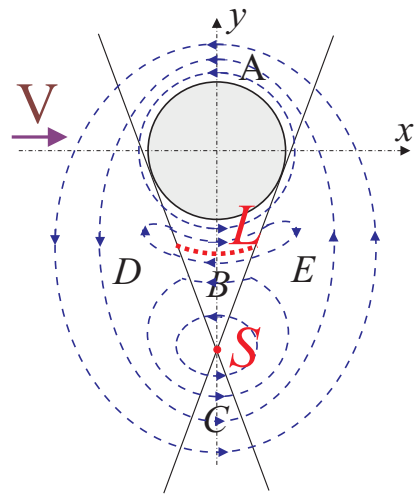}
\end{center}
\caption{\textbf{Fig.3} Schematic distribution of the
superconducting phase gradient $\nabla\phi$ (dashed lines) for a
bound vortex configuration.} \label{Fig:3}
\end{figure}
%

Certainly well above the threshold velocity $V_L$ the gap profile
and stable vortex configuration should be determined from the
self--consistent procedure taking account of contribution of
delocalized excitations. It is natural to expect that the
instability of the bound vortex state and appearance of a freely
moving vortex should be accompanied by the formation of a
continuous anomalous branch connecting the energies below
$-\Delta_0$ and above $+\Delta_0$. It is this branch which
provides necessary conditions for spectral flow through the energy
gap and, thus, for energy dissipation during the vortex escape
from the defect. The obstacles on the way of quasiparticle flow
through the gap could be removed provided the FS segments  would
merge and reconnect to form a cosine curve $\mu(\theta_p)$ typical
for a free vortex with a continuous anomalous spectral branch
(compare solid and dash lines in Fig.~2). This transformation
causes topological changes in FS geometry because new cosine FS is
open and can not be contracted to a point by continuous
transformations. Correspondingly, one can suggest three obvious
scenarios of destruction of the bound vortex configuration:
(i)~first, the FS segments can merge forming an open FS due to
quantum mechanical tunneling; (ii)~second, the spectral flow
through the energy gap can occur due to the impurity scattering
between different Fermi surface segments; (iii)~third, the free
vortex can be formed when the Doppler shift reaches the gap value,
i.e., the current density approaches the depairing one. The first
scenario can be realized due to the Landau-Zener tunneling between
classical orbits and can be understood within the quantum
mechanical picture based on the commutation relation for
canonically conjugated variables $\mu$ and $\theta_p$:
$[\hat{\mu},\hat{\theta}_p]=-i$. Here we use square brackets to
denote a commutator of two operators.
 The efficiency of this tunneling is determined by the ratio of the
distance $2\mu_a$ between the orbits to the angular momentum
uncertainty $\Delta\mu$. The latter value can be estimated from
the uncertainty principle $\triangle\mu \triangle\theta_p \sim 1$
and the equation $\mu=k_\perp V\cos\theta_p/\omega$ describing the
classical orbit for small $\mu$ values. Considering the
trajectories with $\theta_p$ close to $\pm \pi/2$ we find:
$\Delta\mu=k_\perp V\Delta\theta_p/\omega$ and $\Delta\mu \sim
\sqrt{ k_\perp V / \omega}$. Thus, Landau--Zener tunneling results
in the merging of two FS segments provided the condition
$\mu_a<\sqrt{ k_\perp V / \omega}$ is fulfilled. This second
transition in the spectrum accompanied by the change in the Fermi
surface topology occurs at the critical velocity and current given
by the relations:
\begin{equation}
\label{second-cur} V_d  = k_F  \omega a^2  \ ,\,
 j_d = j_c (k_F a^2/\xi)\ln\Lambda
    = k_F a j_L \gg j_L  \ .
\end{equation}
For  $j>j_d$ the FS is described by the Eq.~\ref{orbitt} for all
$\mu$ values (see dashed line in Fig.~2) and corresponds to a
Doppler--shifted spectrum of a free vortex:
$\tilde\varepsilon=-\hbar\omega\mu + \hbar k_F V\cos\theta_p$ for
$|\mu|\ll k_F\xi$. The continuous path for the quasiparticle
spectral flow from the energies below $-\Delta_0$ to the ones
above $+\Delta_0$ is now restored providing conditions for a
dissipative vortex motion.

\begin{figure}
\begin{center}
\includegraphics[width=0.35\textwidth]{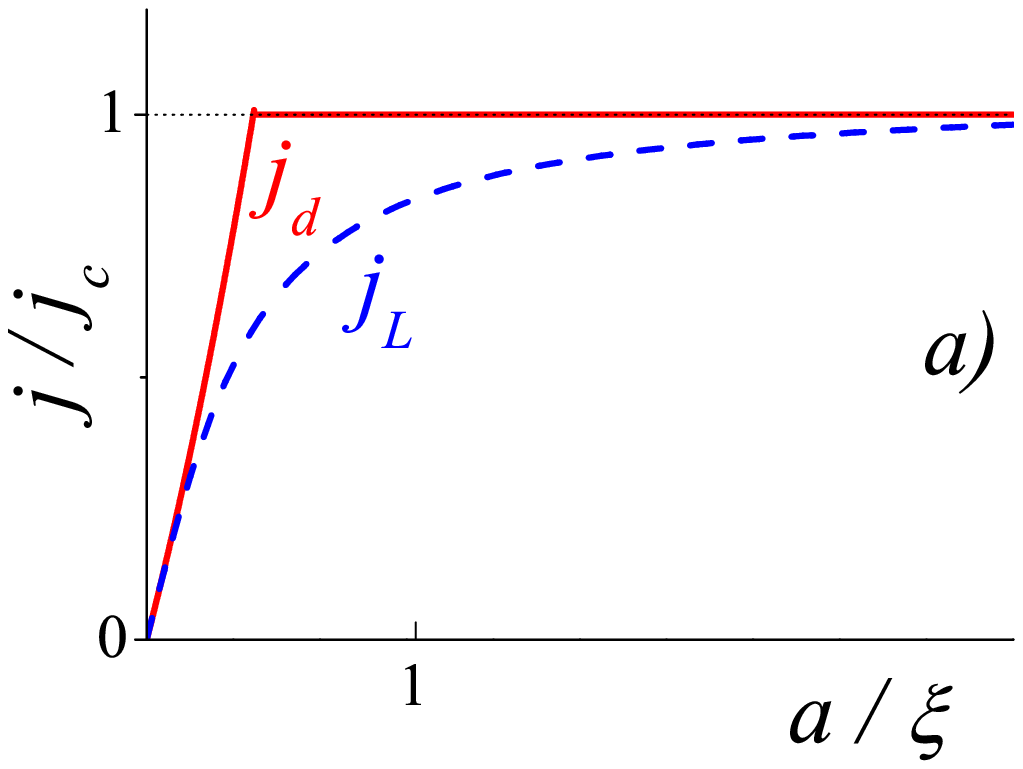}
\includegraphics[width=0.35\textwidth]{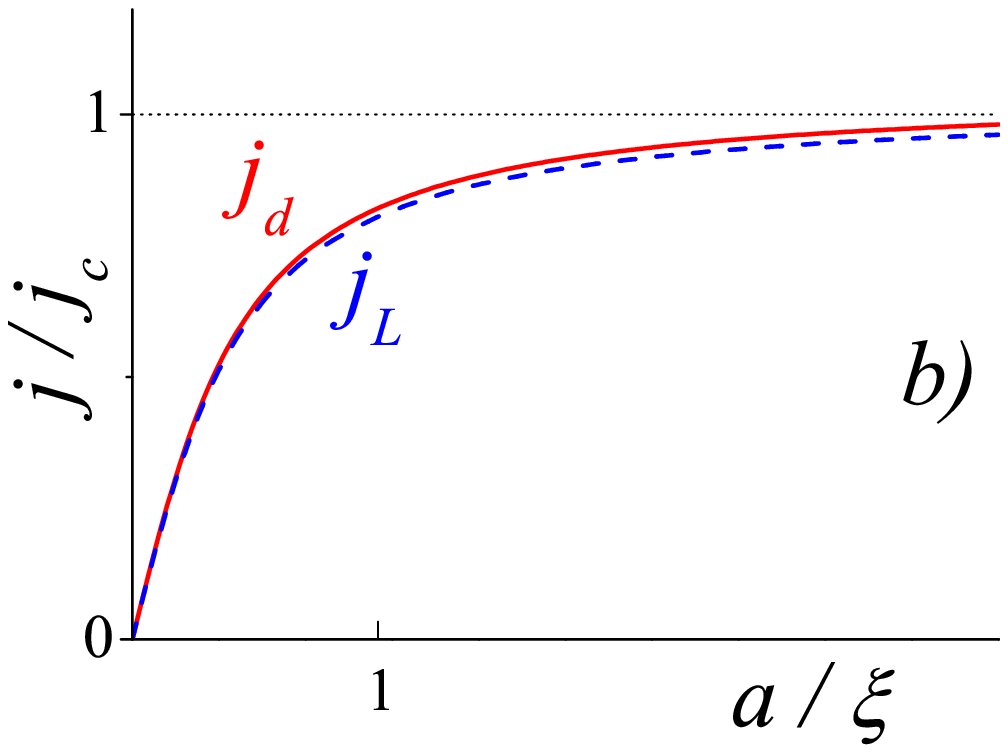}
\end{center}
\caption{\textbf{Fig.4} Schematic dependencies of critical current
densities vs the defect radius $a$ for clean (a) and dirty (b)
limits.} \label{Fig:4}
\end{figure}

Second scenario of the recovering such continuous path through the
gap is associated with impurity scattering. The scattering rate
$1/\tau_s$ between the quasiparticle states at different FS
segments shown in Fig.~2 should be reduced compared to the
impurity scattering rate $1/\tau$ in a normal metal due to the
small size $\delta\theta_p$ of FS segments in the phase plane
${\mu-\theta_p}$: $1/\tau_s\sim \delta\theta_p /\tau$, where
$\cos(\delta\theta_p/2)=\omega a/V$. Provided the scattering rate
$1/\tau_s$ becomes comparable with the minigap
$\Delta_m = k_F a\, \hbar \omega$ for a
pinned vortex the FS segments can be no more considered as
isolated ones. This condition imposes a restriction on the size of
the isolated FS segments
$\delta\theta_p<\delta\theta_{max} = \gamma\Delta_m\tau /\hbar < \pi$
and gives us an estimate for the upper
critical velocity and current which destroy the bound vortex
state:
\begin{equation}\label{second-cur2}
V_d=\frac{V_L}{\cos (\delta\theta_{max}/2)} \ ,\,\,\,
j_d=\frac{j_L}{\cos (\delta\theta_{max}/2)} \ ,
\end{equation}
where the constant $\gamma$ is of the order unity. Note that in
the limit $\gamma\Delta_m\tau /\hbar > \pi$ the impurity
scattering effect is negligible and the depinning current density
is given by the Eq.~\ref{second-cur}. Let us also emphasize that
the critical current of vortex depinning can not exceed the
depairing one. Thus, Eqs.~(\ref{second-cur}, \ref{second-cur2})
are valid only for $j_d < j_c$, i.e., for rather small size of the
defect. In the clean limit, e.g., this restriction on the size
reads: $a \le \sqrt{\xi/k_F} \ll \xi$. For larger defect radii the
depinning current density should saturate at the $j_c$ value. The
resulting schematic dependencies of critical current densities vs
the defect radius $a$ are shown in Fig.~\ref{Fig:4} for clean (a)
and dirty (b) limits. One can see that the unusual bound vortex
state near the defect can be observed only for rather clean
samples with $\Delta_0\tau/\hbar\gg 1$. In the dirty case for
$\Delta_0\tau/\hbar\le 1$ both critical current densities $j_L$
and $j_d$ coincide. The depinning current estimate in this limit
can be also found following the approach developed in
Refs.~\cite{Thuneberg,Thuneberg-JLTP84} for the pinning on
impurities with small scattering cross section.
%
%

To sum up, we have developed a microscopic description of the
two--stage scenario of vortex depinning from small size columnar
defect and predict the formation of stable vortex -- antivortex
configurations bound to the defect at intermediate transport
current densities. The lower critical current density corresponds
to the formation of vortex --antivortex state and should determine
 rf nonlinear response of vortex system at rather small
 transport current amplitudes. The upper critical current density
 $j_d$  corresponds to the destruction of the bound vortex configuration
 around the defect and should determine the depinning transition at
 dc currents.

The authors are thankful to Mike Silaev for many
helpful discussions. This work was supported, in part, by the
Russian Foundation for Basic Research,  Russian Agency of
Education under the Federal Program ``Scientific and educational
personnel of innovative Russia in 2009-2013", and European IRSES
program SIMTECH (contract n.246937).

\end{document}